\documentclass[sn-basic]{sn-jnl}

\usepackage{graphicx}%
\usepackage{multirow}%
\usepackage{amsmath,amssymb,amsfonts}%
\usepackage{amsthm}%
\usepackage{mathrsfs}%
\usepackage[title]{appendix}%
\usepackage{xcolor}%
\usepackage{textcomp}%
\usepackage{manyfoot}%
\usepackage{booktabs}%
\usepackage{algorithm}%
\usepackage{algorithmicx}%
\usepackage{algpseudocode}%
\usepackage{listings}%

\usepackage{mathtools}
\usepackage{multirow}
\usepackage{subcaption}
\usepackage{url}
\usepackage[normalem]{ulem}
\useunder{\uline}{\ul}{}
\usepackage{xcolor}

\usepackage{fancyhdr}
\fancypagestyle{preprint}{
    \fancyhf{}
    
    \fancyhead[C]{\textcolor{gray}{\small \textit{This manuscript is a pre-print.}}}
}

\definecolor{bgtabcolor}{RGB}{247, 210, 173}

\theoremstyle{thmstyleone}%
\theoremstyle{thmstyletwo}%

\theoremstyle{thmstylethree}%

\begin{document}

\title[URecJPQ]{\textit{URecJPQ}: Memory-efficient Multimodal Recommendation Models through RecJPQ in Large-Scale Scenarios}

\author*[1]{\fnm{Giuseppe} \sur{Spillo}}\email{giuseppe.spillo@uniba.it}

\author[2]{\fnm{Zixuan} \sur{Yi}}\email{z.yi.1@research.gla.ac.uk}

\author[3]{\fnm{Aleksandr V.} \sur{Petrov}}\email{aleksandrv@spotify.com}

\author[1]{\fnm{Cataldo} \sur{Musto}}\email{cataldo.musto@uniba.it}

\author[2]{\fnm{Craig} \sur{Macdonald}}\email{craig.macdonald@glasgow.ac.uk}

\author[2]{\fnm{Iadh} \sur{Ounis}}\email{iadh.ounis@glasgow.ac.uk}

\affil[1]{\orgname{University of Bari Aldo Moro}, \city{Bari}, \country{Italy}}

\affil[2]{\orgname{University of Glasgow}, \city{Glasgow}, \country{United Kingdom}}

\affil[3]{\orgname{Spotify}, \city{Glasgow}, \country{United Kingdom}}

\abstract{Training state-of-the-art recommendation models on large-scale industrial datasets can be a challenging task due to the high number of users and items which are typically represented through ID embeddings. Such embeddings typically require a large amount of memory resources, which are not always available. This problem is further exacerbated in multimodal recommendation, in which multimodal item features generally improve recommendation performance, but require more resources to encode. 
In this paper, we introduce \textit{URecJPQ}, a Joint Product Quantization method specifically designed for large-scale and multimodal top-\textit{k} recommendation tasks, in which the vast number of users and items, combined with the available modalities, further increases the memory demands for the computation. 
The core idea is to represent both users and items not as a fully learned, unique embedding, but rather as a concatenation of shared learned sub-embeddings, thereby significantly reducing the total number of trainable parameters. 
Our experiments on three widely-used datasets across different domains (movies, baby and sports products) show that {URecJPQ} can be effectively applied to multimodal recommendation settings. In large scale scenarios, we observe a substantial reduction in checkpoint sizes and the number of trainable parameters (ranging from 86\% to 98\%, and 98\% to 99\%, respectively), with only a marginal decrease in accuracy (8.5\% on recall and 16\% on NDCG, on average), and, in some cases, even performance improvements (up to 85\%), as in the baby products domain. Our codebase is available at \url{https://github.com/giuspillo/urecjpq}.
}


\maketitle
\pagestyle{preprint}

\section{Introduction}

In the last decades, Recommender Systems (RSs) have emerged as an effective technology to tackle the problem of \textit{information overload} and support users in \textit{decision-making} tasks (\cite{Resnick:97}). RSs are often used in e-commerce platforms, in order to match user preferences with potentially relevant items in their catalog.
Many RS approaches have been proposed in the literature, ranging from early \textit{Collaborative Filtering} (\cite{schafer2007collaborative, li2025disentangled}) approaches, to sophisticated ones leveraging {knowledge graphs} (as in \cite{li2026leveraging}) or \textit{multimodal} content to provide personalized suggestions (as in \cite{zhou2023comprehensive, zhang2026mcd4sr, he2025smooth}).

To provide recommendations, RSs typically learn \textit{k}-dimensional embeddings to represent users and items, which are commonly referred to as {\em ID embeddings}. 
ID embeddings are dense vector representations where each user and item's unique identifier is mapped to a continuous vector in a shared latent space (\cite{yuan2023go,zhang2022embedding,he2026latent}).
These vectors can be learned directly from the user-item interaction data (e.g., via matrix factorization or neural collaborative filtering (\cite{xue2017deep,he2017neural})) such that the similarity between a user's embedding and an item's embedding predicts the likelihood of an interaction or the user's preference. 
However, in large-scale recommendation scenarios, the system has to manage a massive number of users and items, learning unique embeddings for each user and for each item. {This introduces our first core challenge: the memory overhead of ID embeddings.}
The embeddings are typically trained on GPUs exploiting backpropagation, which means, on one hand, that the embeddings need to be stored in the GPU, and this requires enough memory. In addition, the backpropagation requires storing additional components, including the gradients, further increasing GPU memory requirements.

\begin{figure}
    \centering
    \includegraphics[width=0.7\linewidth]{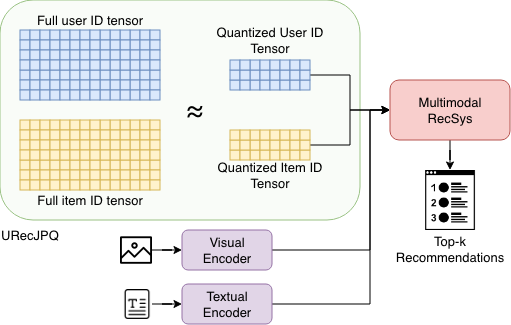}
    \caption{Overview of our URecJPQ method and its application in multimodal recommendation. {The full user and item ID tensors can be approximated and replaced by their quantized counterparts. Then, they are used together with the multimodal features to provide recommendations. }}
    \label{fig:placeholder}
\end{figure}

{For instance, a single recommendation model can contain hundreds of embedding tables with tens of millions of vectors each, leading to state-of-the-art model sizes of multiple terabytes that often exceed the capacity of CPU or GPU memory (\cite{kang2021learning}), while the high-bandwidth memory available on a state-of-the-art GPU is typically only 32--80GB\footnote{\url{https://docs.nvidia.com/deeplearning/performance/recsys-best-practices/index.html}}, making large-scale recommendation models fundamentally memory-bound and starkly illustrating the mismatch between industrial data volume and available GPU memory.}
Similarly, the \textit{Amazon Reviews '23} dataset provided by \cite{hou2024bridging}, which represents only a filtered subset of the much larger global Amazon catalog, contains 54 million users and 48 million items. Storing the full embeddings for these entities would require more than 13 billion parameters, without considering the additional parameters needed for the RS model intermediate layers.

{A second, related challenge arises in multimodal recommendation (\cite{zhou2023mmrec,wang2025structure,wu2025d,ma2025diffkd}), in which} this problem is further exacerbated.
Indeed, multimodal approaches generally improve recommendation performance, as shown by \cite{yi2025enhancing}, but require more trainable parameters to process and encode multimodal data, such as item images or textual descriptions,  {further increasing the computational and memory cost on top of the ID embedding overhead described above}.

A possible solution to the memory requirements for learning full embeddings consists in compressing the embeddings through \textit{quantization}, originaly proposed by \cite{gray1984vector}, in order to reduce the total number of learnable parameters and memory occupancy.
{A third limitation is related to the observation that the} current literature has exploited quantization primarily for sequential recommendation tasks (\cite{mallamaci2025balancing,petrov2024recjpq,rajput2023recommender,zeghidour2021soundstream,xu2026mmq}), in which the aim is to predict the next item users will interact with. 
For example, RecJPQ proposed by \cite{petrov2024recjpq} is a {recent} approach that learns quantized item ID embeddings in an end-to-end manner (instead of post-quantization) for next-item prediction.
However, the use of quantization to improve memory efficiency in traditional top-\textit{k} recommendation scenarios remains overlooked, and even less so in multimodal recommendation.

In this paper, we address this gap by extending RecJPQ to the multimodal and top-\textit{k} recommendation setting for large-scale catalogs. 
{We note that the primary novelty of this work lies in successfully applying and empirically validating this extension in a previously unexplored setting.}
In doing so, this paper makes the following contributions: 
\begin{itemize}
    \item We successfully extend RecJPQ to the top-\textit{k} recommendation scenario, by quantizing not only item ID embeddings as in sequential recommendation, but also user ID embeddings -- we call this extension \textbf{URecJPQ}. {This extension is particularly relevant for the industrial top-\textit{k} scenarios in which both user and item embedding tables can independently become memory bottlenecks at scale.}
    \item We apply URecJPQ to multimodal recommendation in large-scale catalog scenarios, thus drastically reducing the number of trainable parameters. {This enables training and deploying multimodal recommendation models under the limited GPU memory budgets commonly faced in industrial settings, without requiring specialized or costly hardware.}
    \item We quantitatively analyze the trade-off between the decrease of the total number of trainable model parameters (and model checkpoint sizes) and the recommendation performance. {This analysis provides practitioners with concrete evidence to guide deployment decisions, showing that substantial memory savings can be achieved with only marginal, and in some cases negligible, impact on recommendation quality.}
\end{itemize}

This paper is structured as follows: in Section~\ref{sec:rel_works}, we analyze related works, while in Section~\ref{sec:methodology} we describe our methodology. In Section~\ref{sec:exp_setting}, we describe our experimental settings and in Section~\ref{sec:discussion} we discuss our results from a quantitative perspective, while in Section~\ref{sec:qualitative_analysis} we provide a qualitative analysis. Finally, in Section~\ref{sec:conclusions}, we draw our conclusions and sketch future works.

\section{Related Work}\label{sec:rel_works}

In this section, we {first introduce the state-of-the-art of the quantization problem, then we} review related papers in the recommendation domain that leverage quantization approaches.

\subsection{Product Quantization}

In order to reduce the memory occupancy and the number of trainable parameters, a possible solution is the Product Quantization (PQ) approach (\cite{gray1984vector,jegou2010product}), which is commonly used to compress collections of embeddings: its core idea is to split each embedding into smaller, non-overlapping sub-embeddings, and then independently quantize each sub-embedding by mapping it to the index of its closest centroid from a learned codebook (i.e., sub-IDs).
As a result, the full embedding can be approximately reconstructed by concatenating its sub-embeddings, and used in any downstream task (including recommendation). 
\cite{jegou2010product} apply PQ principle to the approximate nearest neighbor search task showing increased efficiency, better performance, and excellent scalability to even large datasets.

However, PQ requires enough memory to train the original full embeddings before compressing them (as shown by \cite{petrov2024recjpq}), and, since quantization is not differentiable, it is impossible to jointly compress the embeddings and perform downstream tasks in an end-to-end manner. 

This limitation is addressed by the Joint Product Quantization method introduced in \cite{zhan2021jointly} (JPQ), a PQ approach that assigns the sub-IDs before training the backbone model for the {document retrieval} task.
The sub-ID assignment is the only non-differentiable operation in PQ, therefore, if the assignments occur before the model training, it can be trained in an end-to-end manner without learning the full embeddings. Then, when the full embeddings are needed, they are reconstructed by concatenating the sub-embeddings. 
In other words, JPQ can replace the embedding tensors of any backbone model without requiring other changes to the model itself. 

\subsection{Quantization in Recommendation}

Quantization approaches have also been exploited in recommendation scenarios, {as discussed by \cite{liu2024vector}}. 
For example, \cite{lian2020lightrec} proposed LightRec, a recommendation model that compress only item ID embeddings through additive quantization (\cite{babenko2014additive}), and uses the Deep Semantic Similarity Model (porposed by \cite{huang2013learning}) as the backbone model together with a knowledge distillation component. However, it learns full item ID embeddings, and it is unclear whether the quantization can be used outside the Deep Semantic Similarity Model.
{\cite{xiao2026beyond} propose Dense Residual Quantization (DRQ), a framework based on Residual Quantization that preserves semantic integrity throughout the quantization process. This is achieved by reincorporating the original content embedding at each quantization layer by concatenating it with the corresponding residual vector. \cite{xu2026mmq} propose Multimodal Mixture-of-Quantization, a two-stage tokenizer that first learns semantic IDs via a shared-specific multi-expert quantization architecture to capture cross-modal synergy and modality-specific uniqueness, and then fine-tunes them with a behavior-aware objective to bridge the semantic-behavioral gap for recommendation.}

\cite{rajput2023recommender} proposed TIGER, an extension of the Residual-Quantized Variational Autoencoder (RQ-VAE \cite{zeghidour2021soundstream}) originally developed in the computer vision field. TIGER is a recommendation model inspired by generative retrieval methods (\cite{tay2022transformer}) and uses textual information to generate item representations.
Instead of relying solely on ID-based embeddings, TIGER uses textual side information during quantization to better capture item hierarchies, thereby improving next-item prediction in sequential recommendation. In principle, TIGER can exploit other modalities (e.g., visual features) rather than only textual one, but this has never been investigated in the literature.

\cite{petrov2024recjpq} proposed RecJPQ, an adaption of the JPQ method to the sequential recommendation scenario. 
Unlike TIGER, RecJPQ does not use residual quantization or learn hierarchical or semantic embeddings. 
In RecJPQ, instead of learning full and unique item ID embeddings, the authors applied the JPQ principles to a number of backbone sequential models (BERT4Rec proposed by \cite{sun2019bert4rec}, GRU proposed by \cite{hidasi2015session}, SASRec proposed by \cite{hidasi2015session}) to significantly reduce the sizes of the trained models while maintaining comparable -- or even improved -- recommendation performance. {A graphical illustration of RecJPQ is provided in Figure~\ref{fig:recjpq}.}
Training shared sub-embeddings dramatically reduces the number of trainable parameters and the gradients required for backpropagation, thereby lowering memory usage and enabling training on modern GPUs -- even for large-scale catalogs.

However, none of the aforementioned approaches (except LightRec) have investigated the impact of quantization in either traditional top-\textit{k} or multimodal recommendation, where both user and item ID embeddings are required to compute dot products and, consequently, recommendation scores.
In contrast, existing methods such as TIGER and RecJPQ focus solely on quantizing item embeddings. Since we are also interested in user ID embeddings, and TIGER learns semantic item embeddings, our work builds on RecJPQ. 

To sum up, the novelty of our URecJPQ is twofold: first, it is the first to study quantization for both user and item ID embeddings in the context of top-k recommendation; second, it is the first to investigate quantization techniques in large-scale and multimodal recommendation settings.

\section{Methodology}\label{sec:methodology}

We now describe the original RecJPQ introduced by \cite{petrov2024recjpq} approach, which serves as a basis for our paper, and introduce URecJPQ, our extension for top-\textit{k} recommendation.

\subsection{Basics of RecJPQ}

\begin{figure}
    \centering
    \includegraphics[width=0.8\textwidth]{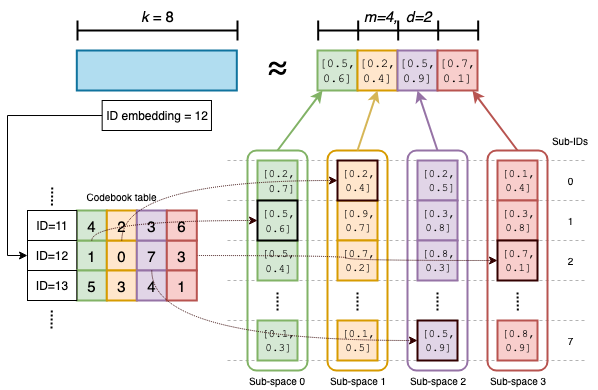}
    \caption{Embedding quantization in RecJPQ: an 8-dimensional embedding is approximated by the concatenation of 4 sub-embeddings, each 2-dimensional.}
    \label{fig:recjpq}
\end{figure}

As mentioned above, RecJPQ allows an end-to-end training of sequential recommendation models with quantized item embeddings.
In particular, given a collection of $k$-dimensional embeddings $\{v_1, v_2, ..., v_n\}$ {(for example, the items in a catalog)}, RecJPQ learns a number of $d$-dimensional sub-embeddings such that $d \ll k$ and $d=\frac{k}{m}$, with $m$ being the number of sub-spaces considered for the quantization. The number of the sub-embeddings in each sub-space, denoted as the codebook length $l$, is typically a power $p$ of $256$, so that the codes can be stored in a $p$-byte integer structure. 
Following \cite{petrov2024recjpq} and \cite{zhan2021jointly}, by setting $p=1$, each code can be represented by using a single byte. As a result, each sub-space has $256$ sub-embeddings.

In addition, RecJPQ supports multiple strategies for code assignment. The simplest strategy, \textbf{random}, assigns each item $m$ random sub-IDs (from $0$ to $255$), one per subspace. Its main limitation is that it may assign different sub-embeddings to similar items -- or the same sub-embeddings to diverse items -- risking a loss of semantic coherence.

To preserve semantic consistency across items, \textbf{Truncated SVD} (\cite{hansen1987truncated}) offers an alternative code assignment strategy. Specifically, performing SVD with $m$ latent factors on the interaction matrix $Y$ provides low-dimensional user and item embeddings, such that similar users (items) are mapped to similar embeddings. Formally, let $Y \approx U \cdot \Sigma \cdot I^T$, where $U$ and $I$ are the user and item embeddings, and $\Sigma$ is a diagonal matrix of singular values.

The resulting item embeddings $I$ are normalized, and a small random noise is added to avoid identical embeddings for items with identical interaction histories.
Each of the $m$ dimensions of the normalized embeddings is then discretized into $b$ quantiles, each containing roughly the same number of items. These quantile bins define the $m$ sub-IDs for each item -- one per subspace.

Figure~\ref{fig:recjpq} {provides an example of this approach. Instead of learning isolated $8$-dimensional embeddings, one per item, only $4$ sub-spaces are learnt, each composed of smaller $2$-dimensional sub-embeddings. A codebook table is used to map each full embedding with a sequence of sub-embeddings (though sub-IDs), that can be used to reconstruct the full embedding when needed.
} 

\subsection{URecJPQ}

In \cite{petrov2024recjpq}, the process described above is applied exclusively to item embeddings. This design choice aligns with the requirements of \textit{sequential} recommendation, where users are typically represented by their interaction sequences rather than explicit ID embeddings. 

{In contrast, our proposed URecJPQ method extends this quantization framework to include user ID embeddings alongside item ID embeddings. This dual quantization is essential for the top-$k$ recommendation scenario, where explicit vector representations for both users and items are required to compute recommendation scores. By quantizing both user and item ID embeddings, URecJPQ maintains the necessary interaction logic while achieving the memory efficiency required for large-scale catalogs.}

{
Also in this case, given $k$-dimensional original embeddings, we aim to learn $d$-dimensional sub-embeddings such that $d \ll k$ and $d=\frac{k}{m}$, with $m$ being the number of sub-spaces considered for the quantization, and $l$ the codebook lenght (number of embeddings in each sub-space). To this aim, URecJPQ uses two code assignment stragies to builds the codebook tables for users and items, random and TruncatedSVD, similarly to RecJPQ.
} 

In the random code assignment strategy, random sub-IDs are assigned to both users and items, resulting in two codebook tables that are populated at random. 
In the Truncated SVD strategy, URecJPQ preserves semantic interaction patterns by first decomposing the user-item interaction matrix $Y \approx U \cdot \Sigma \cdot I^T$, where $U$ and $I$ represent the latent factors for users and items, respectively. Item embeddings are stored in $I^T$, while user embeddings are stored in $U$. 
To map these continuous factors to discrete codes, we set the number of SVD components equal to the number of subspaces $m$, such that each latent dimension $i \in \{1, \dots, m\}$ corresponds to a specific subspace. 
For each dimension, we apply quantile discretization using $b$ bins. 
This process ensures that users and items are distributed evenly across the available sub-IDs in each subspace, effectively transforming the most significant global interaction signals into structured, discrete identifiers. 
To prevent collisions between entities with identical interaction histories, a small random noise ($\epsilon \approx 10^{-5}$) is added to the embeddings prior to discretization.

A distinctive trait of RecJPQ is its ability to support {the} end-to-end training of the sequential recommendation models into which it is integrated. URecJPQ inherits this property, making it applicable to \textit{any} backbone model that explicitly uses user and item ID embeddings to provide top-\textit{k} recommendation (including most of multimodal models): this improves their efficiency in terms of memory occupancy and number of trainable parameters.

\begin{figure}
  \centering
  \begin{subfigure}[b]{0.6\textwidth}
    \centering
    \includegraphics[width=\textwidth]{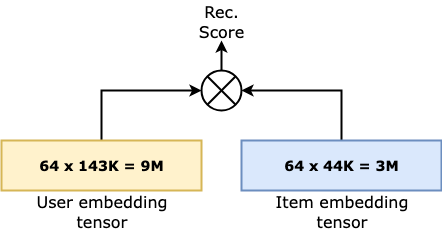}
    \caption{Traditional BPR model. The number of trainable parameters depends on the number of users (143K in the example), items (44K) and embedding size (64){, for a total of 12M trainable parameters}.}      
    \label{fig:bpr_example}
  \end{subfigure}
  \hspace{2cm}
  \begin{subfigure}[b]{0.6\textwidth}
    \centering
    \includegraphics[width=\textwidth]{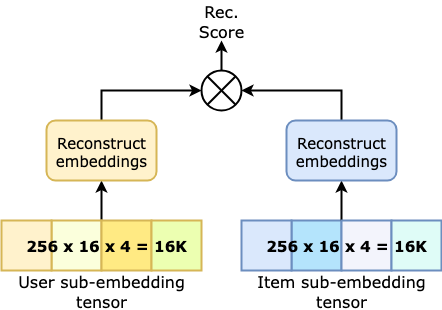}
    \caption{URecJPQ applied to the BPR model. The number of trainable parameters depends on the number of sub-IDs per space (256 in the example), the number of spaces (16), and the original sub-embedding size, computed as the ratio between the original embedding size (64) and the number of subspaces(4){, for a total of 32K trainable parameters}.}
    \label{fig:urecjpq_bpr_example}
  \end{subfigure}
  \caption{Comparison between the number of trainable parameters of the BPR model on the Baby23 dataset, for 143K users and 44K items. The vanilla version has 12M total trainable parameters, while URecJPQ has only 32K{, with a reduction of 99.73\%}.}
  \label{fig:urecjpq_vs_normal}
\end{figure}

{To illustrate the parameter efficiency of URecJPQ, we consider the BPR model (\cite{rendle2012bpr}), whose trainable parameters consist solely of user and item ID embeddings.
In Figure~\ref{fig:urecjpq_vs_normal}, we compare} BPR \cite{rendle2012bpr} and its quantized version URecJPQ-BPR Baby 23 dataset. 

In vanilla BPR (Fig.~\ref{fig:bpr_example}), the total number of trainable parameters is computed as $(\#users + \#items) * k$, with $k$ being the embedding size ($64$ in the example). 
{This results in a total of $(143K+44K)*64=12M$ parameters.}
In URecJPQ-BPR (Fig.~\ref{fig:urecjpq_bpr_example}), the total number of trainable parameters becomes $256 * m * d$, with $m$ being the number of sub-spaces and $d=\frac{k}{m}$ the sub-embedding size. 
{This results in a total of $256*16*4=16K$ parameters for user embeddings, and $256*16*4=16K$ for item embeddings, for a total of $32K$ parameters.}

{From this example, we can see that
URecJPQ dramatically decreases the number of trainable parameters, from 12M to 32K for BPR (99.7\%). When considering VBPR (\cite{he2016vbpr}), the multimodal variant of BPR, this decrease is even exacerbated, as the number of trainable parameter decreases from 21M to 170K.}

\subsection{Summary of URecJPQ}

URecJPQ is an extension of the original RecJPQ applied in a top-\textit{k} recommendation scenario. While RecJPQ only quantizes item embeddings, URecJPQ extends the approach by also quantizing user embeddings. This is necessary since top-\textit{k} recommendation requires both user and item ID embeddings, while sequential recommendation only involves item embeddings. 
In this way, URecJPQ can be applied to \textit{any} backbone recommendation model, thus replacing user and item tensors in the model design.

Given its flexibility, we argue that this paper represents an important contribution to the field, as {URecJPQ can be integrated in any recommendation model that leanrs user and item ID embeddings}, thus improving their memory efficiency and total number of trainable parameters.
Moreover, parameter sharing across users and items also acts as an effective regularization mechanism to reduce overfitting (as shown by \cite{petrov2024recjpq}), and better clusters users and item. We leave the investigation of such aspects for future work.

To show its effectiveness, we apply URecJPQ in large-scale and multimodal recommendation scenarios.

With respect to the other works in the area, such as LightRec, TIGER, {DRQ, MMQ,} and RecJPQ, our URecJPQ approach enables an end-to-end training of quantized both user and item ID embeddings, rather than only item embeddings (see Table~\ref{tab:comparison}). Moreover, it is designed for the traditional top-\textit{k} recommendation setting, instead of sequential recommendation, as is the case for all other considered methods. Because of these aspects, it can be integrated in any backbone recommendation model, including multimodal, in contrast to TIGER {and DRQ}, which only partially support side information from a single modality, LightRec, which does not support multimodality at all, {and MMQ, which, despite supporting multimodality, remains limited to the sequential recommendation setting and quantizes item embeddings only.} As shown in Table~\ref{tab:comparison}, none of the existing approaches jointly quantize user and item embeddings for multimodal top-\textit{k} recommendation, a gap that URecJPQ is the first to address.

\begin{table}
\centering
\renewcommand{\arraystretch}{1.2} 
\resizebox{0.8\textwidth}{!}{
\begin{tabular}{ccccc}
\hline \textsc{} & \textsc{End-to-end} & \textsc{Multimodal} & \textsc{Quantization} & \textsc{Type} \\ \hline 
LightRec & \checkmark & X & Item & Top-k \\
TIGER & \checkmark & {partially} & Item & Sequential \\ 
DRQ & \checkmark & {partially} & Item & Sequential \\
MMQ & \checkmark & \checkmark & Item & Sequential \\
RecJPQ & \checkmark & X & Item & Sequential \\ \hline 
URecJPQ & \checkmark & \checkmark & User + Item & Top-k \\ \hline 
\end{tabular}
}
\caption{Comparison of our URecJPQ with other similar quantization approaches, highlighting the novelty and the differences of our proposed model with respect to the state of the art.}
\label{tab:comparison}
\end{table}

\section{Experimental Design}\label{sec:exp_setting}

We conduct our experiments to answer the following research questions: 
\begin{itemize}
    \item \noindent\textbf{RQ1:} How does URecJPQ affect the performance of state-of-the-art multimodal recommendation models?

    \item \noindent\textbf{RQ2:} How do different number of sub-spaces affect the overall performance?

    \item \noindent\textbf{RQ3:} How do different code assignment strategies (random, SVD) affect overall performance?
\end{itemize}

\begin{table}[t]
\centering
\resizebox{0.8\textwidth}{!}{
\renewcommand{\arraystretch}{1.2} 
\begin{tabular}{l|ccc||cc}
\hline
 & \textsc{ML1M} & \textsc{Baby 23} & \textsc{Sports 23} & \textsc{Baby 14} & \textsc{Sports 14} \\
\hline 
\# users        & 6,040     & 143,602   & 338,058   & 19,445    & 35,598 \\
\# items        & 2,980     & 44,340    & 165,334   & 7,037     & 18,287 \\
\# interactions & 942,703   & 1,178,208 & 2,852,781 & 160,522   & 295,366 \\ 
Density         & 0.94763\% & 0.9998\%  & 0.9999\%  & 0.9998\%  & 0.9995\% \\ 
Modalities      & T, V      & T, V      & T, V      & --        & -- \\
\hline
\end{tabular}
}
\caption{Statistics of the datasets used in our experiments (on the left), compared to previous and smaller versions of the same datasets (on the right).}
\label{tab:dataset_stats}
\end{table}

\noindent\textbf{Datasets.} 
We evaluate our methodology {on three datasets, of which two are} large-catalog datasets provided with multimodal information. 
In particular, we use \textit{Baby23} and \textit{Sports23} from the Amazon Review 2023\footnote{\url{https://amazon-reviews-2023.github.io}} datasets provided by \cite{hou2024bridging}. 
{In addition, we consider the MovieLens-1M (\textit{ML1M}) dataset, provided by \cite{spillo2025see}.
ML1M already has multimodal data, so no additional processing is needed. On the other hand, to process Baby23 and Sports23 these datasets,} we follow common approach in the literature, as in \cite{yi2025enhancing,yi2022multi}: first, we apply a \textit{core-10} filtering, to keep only users and items with at least 10 interactions. Then we filter out all items for which multimodal information is unavailable. 
As multimodal information, we use images and textual descriptions, since these are the available modalities in the datasets, but we note that our method can be applied to any number of modalities. {We use MiniLM (\cite{wang2020minilm}) and ViT (\cite{dosovitskiy2020vit}) models to encode textual and visual data, respectively. The same encoders have been used for the ML1M, as described in \cite{spillo2025see}.}
{We rely on pre-trained frozen multimodal features to align with the current research (\cite{yi2022multi,zhou2023tale, zhang2021mining,wei2019mmgcn,tao2022self}). This is justified by the fact that, if the encoders were instead trained end-to-end, they alone would introduce tens to hundreds of millions of additional trainable parameters, dwarfing the ID embedding tensors and confounding the memory savings URecJPQ achieves.}

The statistics of the post-processed datasets are reported in Table~\ref{tab:dataset_stats}. The scripts to download the original datasets and process them are provided in our repository\footnote{\url{https://github.com/giuspillo/urecjpq}}, along with the post-processed version used in our experiments, and the source code to reproduce our results.
We highlight the scale of the datasets we use: previous versions of these datasets, containing data up to 2014, were considerably smaller as depicted by the right-hand side of the table. {Indeed, the 2023 versions represent a significant expansion over the 2014 iterations, containing over nine times the interactions, approximately seven times the number of users, and nine times the number of items. This growth in the cardinality shows the large-scale of the datasets we consider.}
A deeper discussion is provided by \cite{hou2024bridging}.

\noindent\textbf{Recommendation models. }
We apply our methodology to four state-of-the-art
recommendation models commonly used in collaborative filtering and multimodal recommendation, evaluating them in a top-\textit{k} recommendation setting. Note that each of these models needs to learn user and item ID embeddings, thus their memory efficiency can be improved with our URecJPQ, as previously discussed in Section~\ref{sec:methodology}. In particular, we use the following models:
\begin{itemize}
    \item \textbf{BPR} (Bayesian Personalized Ranking), proposed by \cite{rendle2012bpr}: a top-\textit{k} pairwise ranking model that optimizes personalized recommendations by maximizing the difference in predicted scores between a user's preferred and non-preferred items. 
    \item \textbf{VBPR} (Visual Bayesian Personalized Ranking), proposed by \cite{he2016vbpr}: the adaptation of the original BPR model for handling multimodal features, which are concatenated to the item ID embeddings.
    \item \textbf{SLMRec} (Self-supervised Learning for Multimedia Recommendation), proposed by \cite{tao2022self}: a self-supervised graph learning model that uses the available modalities to generate supervised signals, which are then used as components of a contrastive loss, ultimately aligning the various modality spaces.
    \item \textbf{MMGCL} (Multi-Modal Graph Contrastive Learning), proposed by \cite{yi2022multi}: a self-supervised multi-task graph learning model that performs edge dropout and modality masking to better model user preferences, learning the correlation between different modalities. It uses both a self-supervised and a BPR loss. 
\end{itemize}

\noindent\textbf{Implementation details. }
Moreover, to guarantee and simplify the replicability of our study, we implement our methodology within the MMRec released by \cite{zhou2023mmrec} framework, which already provides the implementations of all the aforementioned recommendation models. 

\noindent\textbf{Evaluation protocol. }
Following the literature in the field (\cite{zhang2021mining,zhou2023comprehensive,yi2025enhancing}), we split our datasets into training, validation and testing data with an 80:10:10 ratio.
We evaluate the performance using typical top-\textit{k} recommendation metrics, namely Recall@20 and NDCG@20. 
Moreover, since our focus is to investigate how URecJPQ can be applied in large-scale scenarios, we also use as metrics the checkpoint sizes of the trained models and the number of trainable parameters of each model.
As the validation metric, we employ Recall@20 with early stopping after 10 consecutive epochs without improvement, {following MMRec as in \cite{zhou2023mmrec}}. Finally, we apply the Two One-Sided Tests (TOST) to assess whether the quantized models are statistically equivalent to the vanilla models. We set the equivalence margin $\epsilon=0.1$, meaning that differences in performances smaller than $0.1$ are considered negligible.

\noindent\textbf{Hyperparameters. }
We conduct a grid search over the quantization parameters of URecJPQ and the backbone model-specific parameters.
In particular, we learn, for the vanilla version of the considered models, full embeddings of sizes $k=\{64, 128, 256\}$. Then, we apply the quantization with the codebook length fixed to $l=1$, and explore different number of sub-spaces, in the range of $m=\{8, 16, 32, 64, 128\}$, as well as two code assignment strategies, namely, \textit{random} and \textit{SVD}.
Due to space reasons, we are unable to provide the hyperparameter values specific to the backbone models here. Instead, we direct readers to our aforementioned repository for full details. Notably, we optimized these parameters for both the vanilla and quantized versions of the models.

\noindent\textbf{System Configuration.} The MMRec framework is built on PyTorch. We use version 2.6 along with CUDA driver version 12.4. Our experiments are conducted on a platform equipped with an A5000 GPU and 24 GB of RAM.

\section{Results Discussion}\label{sec:discussion}

\begin{figure*}
    \centering

    \begin{subfigure}[b]{0.48\textwidth}
        \centering
        \includegraphics[width=\textwidth]{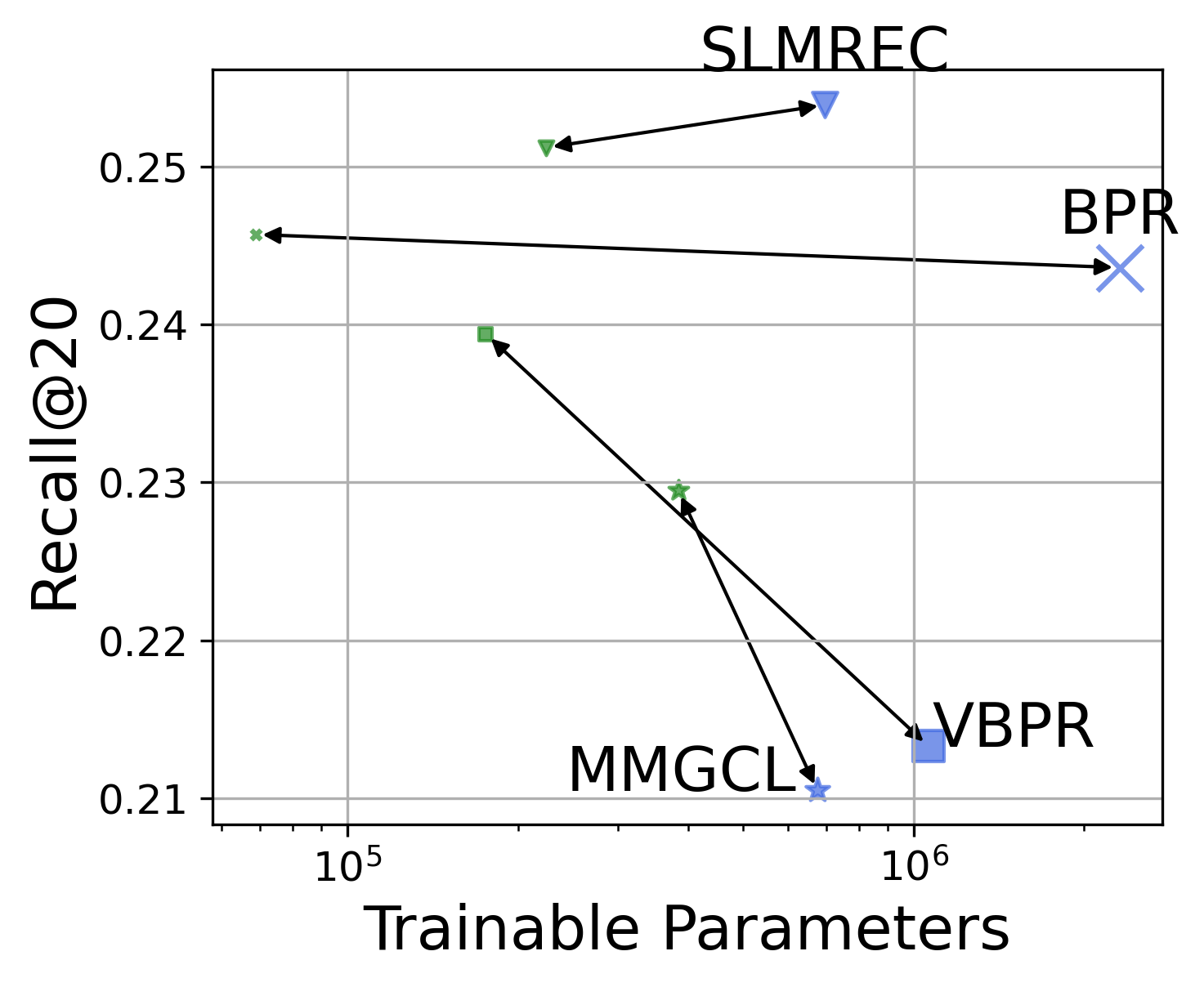}
        \caption{Recall on ML1M}
        \label{fig:rq1_ml1m_recall}
    \end{subfigure}
    \hfill
    \begin{subfigure}[b]{0.48\textwidth}
        \centering
        \includegraphics[width=\textwidth]{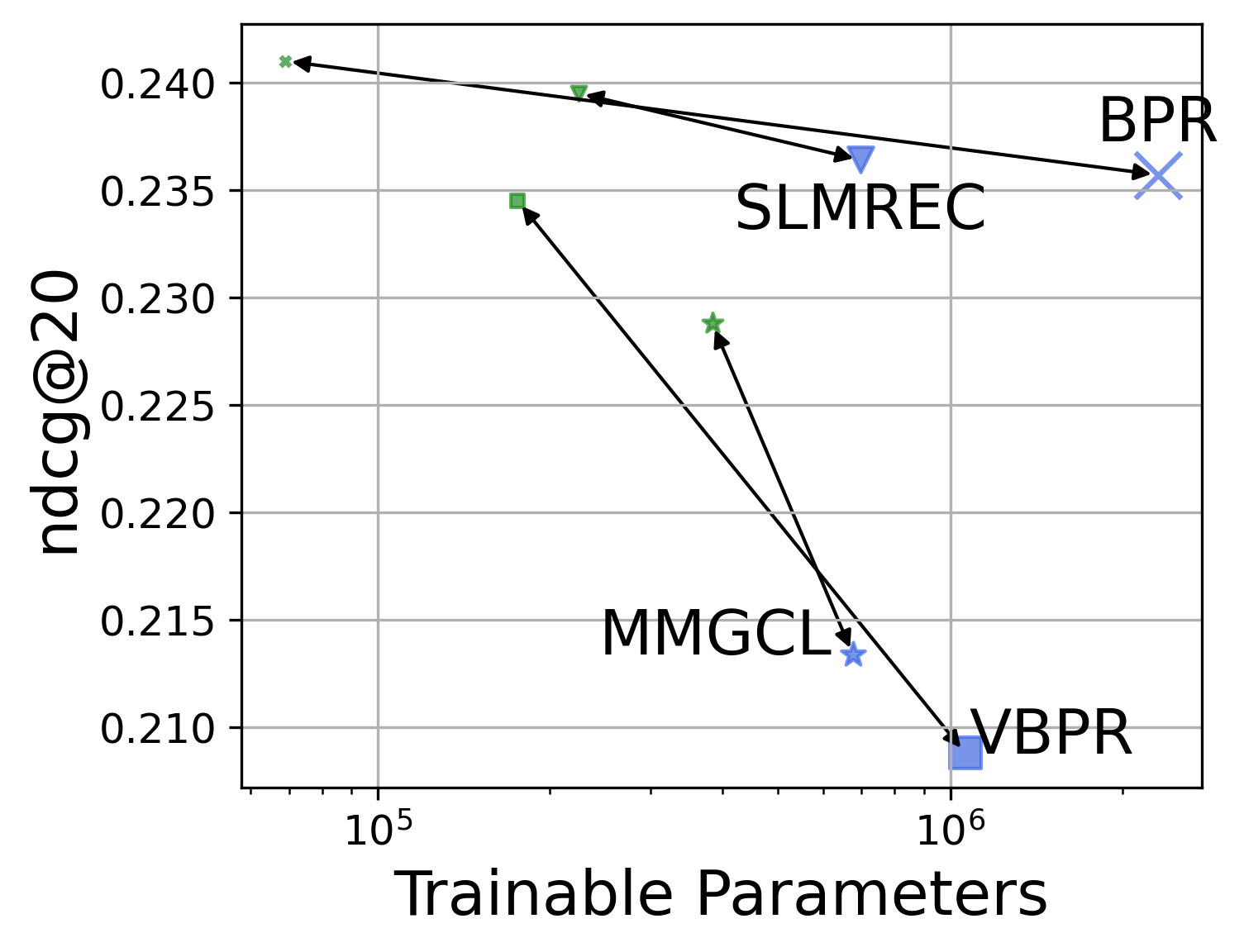}
        \caption{NDCG on ML1M}
        \label{fig:rq1_ml1m_ndcg}
    \end{subfigure}

    \vspace{1em} 

    \begin{subfigure}[b]{0.48\textwidth}
        \centering
        \includegraphics[width=\textwidth]{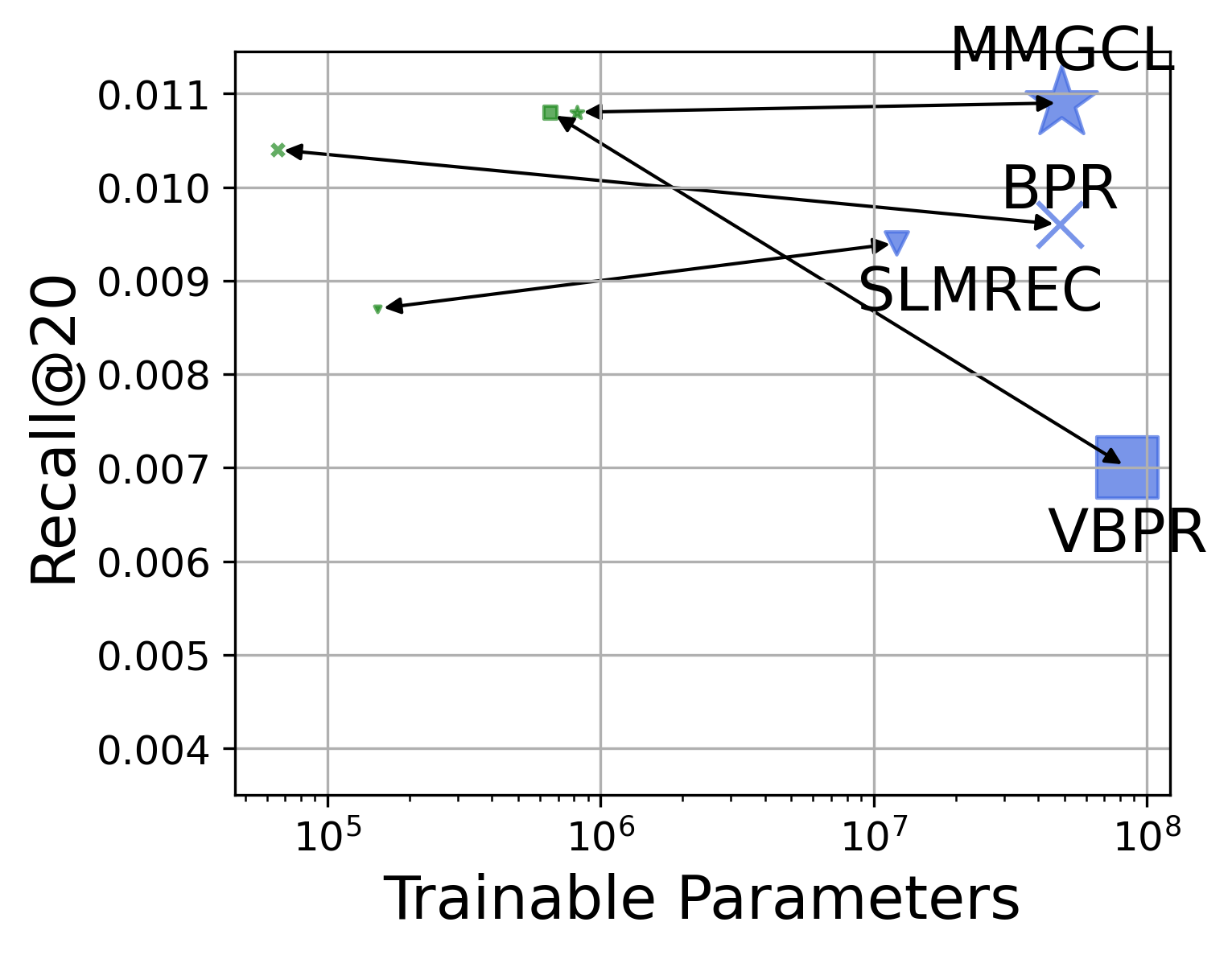}
        \caption{Recall on Baby23}
        \label{fig:rq1_baby_recall}
    \end{subfigure}
    \hfill
    \begin{subfigure}[b]{0.48\textwidth}
        \centering
        \includegraphics[width=\textwidth]{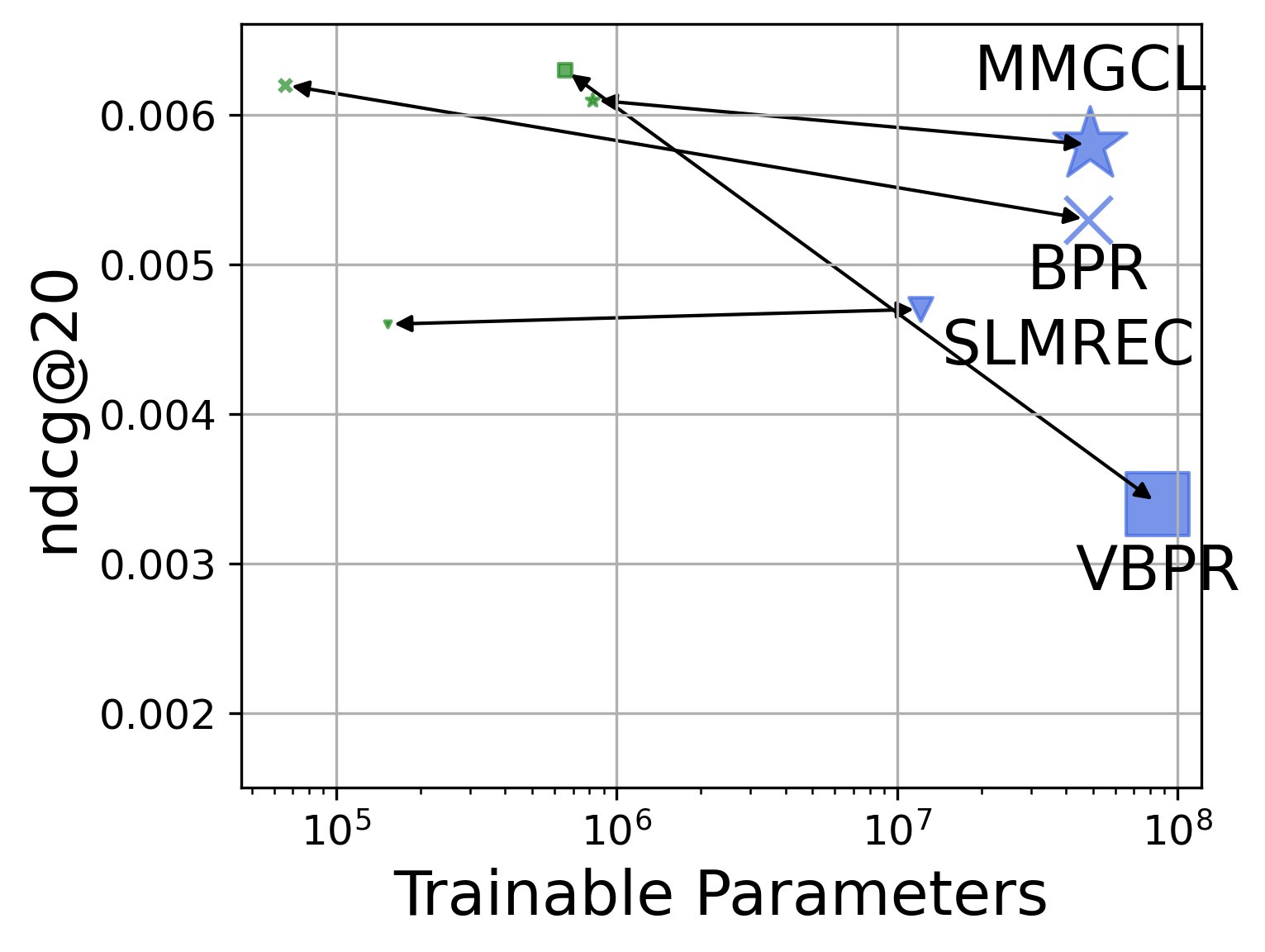}
        \caption{NDCG on Baby23}
        \label{fig:rq1_baby_ndcg}
    \end{subfigure}

    \vspace{1em} 

    \begin{subfigure}[b]{0.48\textwidth}
        \centering
        \includegraphics[width=\textwidth]{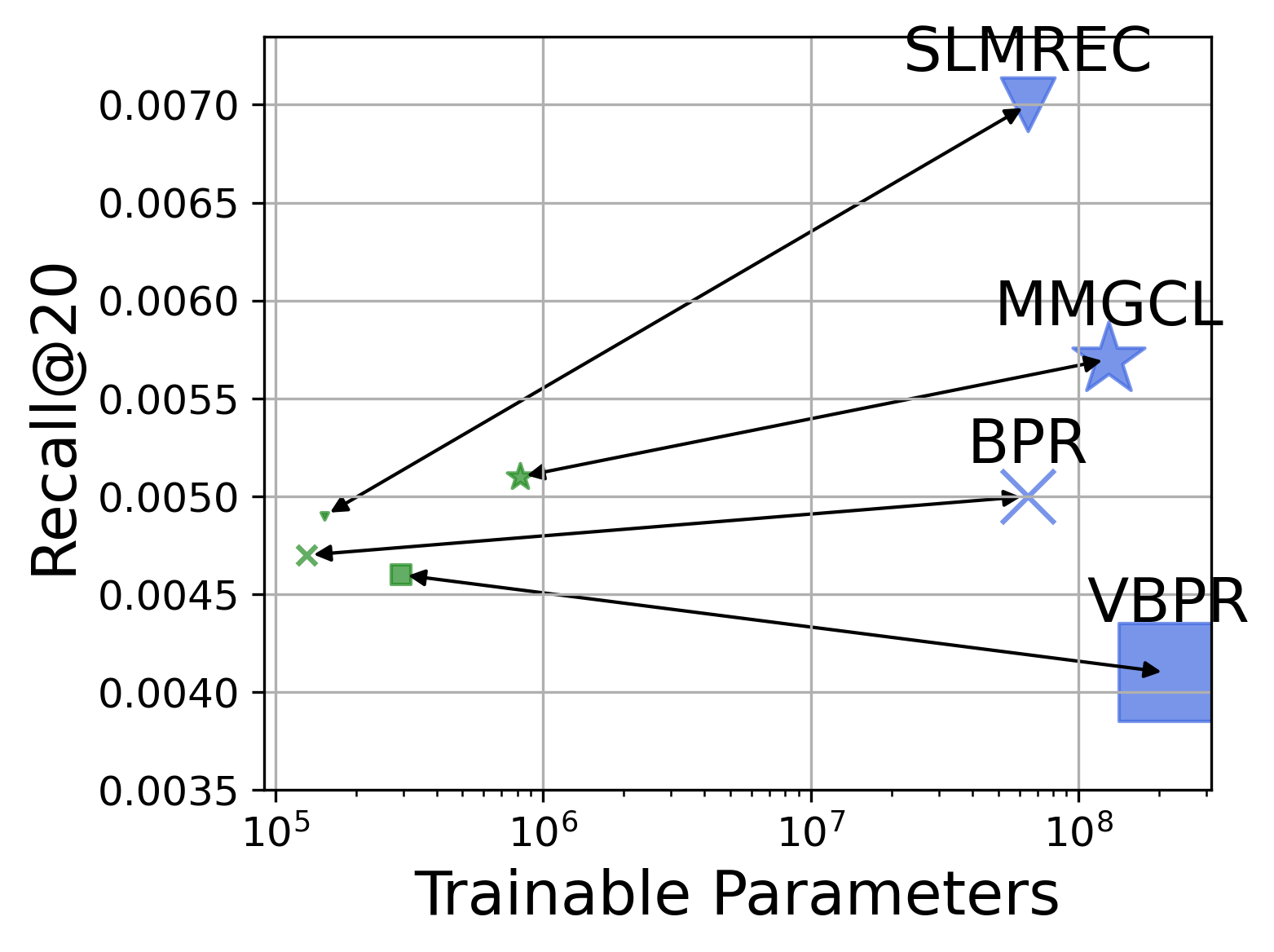}
        \caption{Recall on Sports23}
        \label{fig:rq1_sports_recall}
    \end{subfigure}
    \hfill
    \begin{subfigure}[b]{0.48\textwidth}
        \centering
        \includegraphics[width=\textwidth]{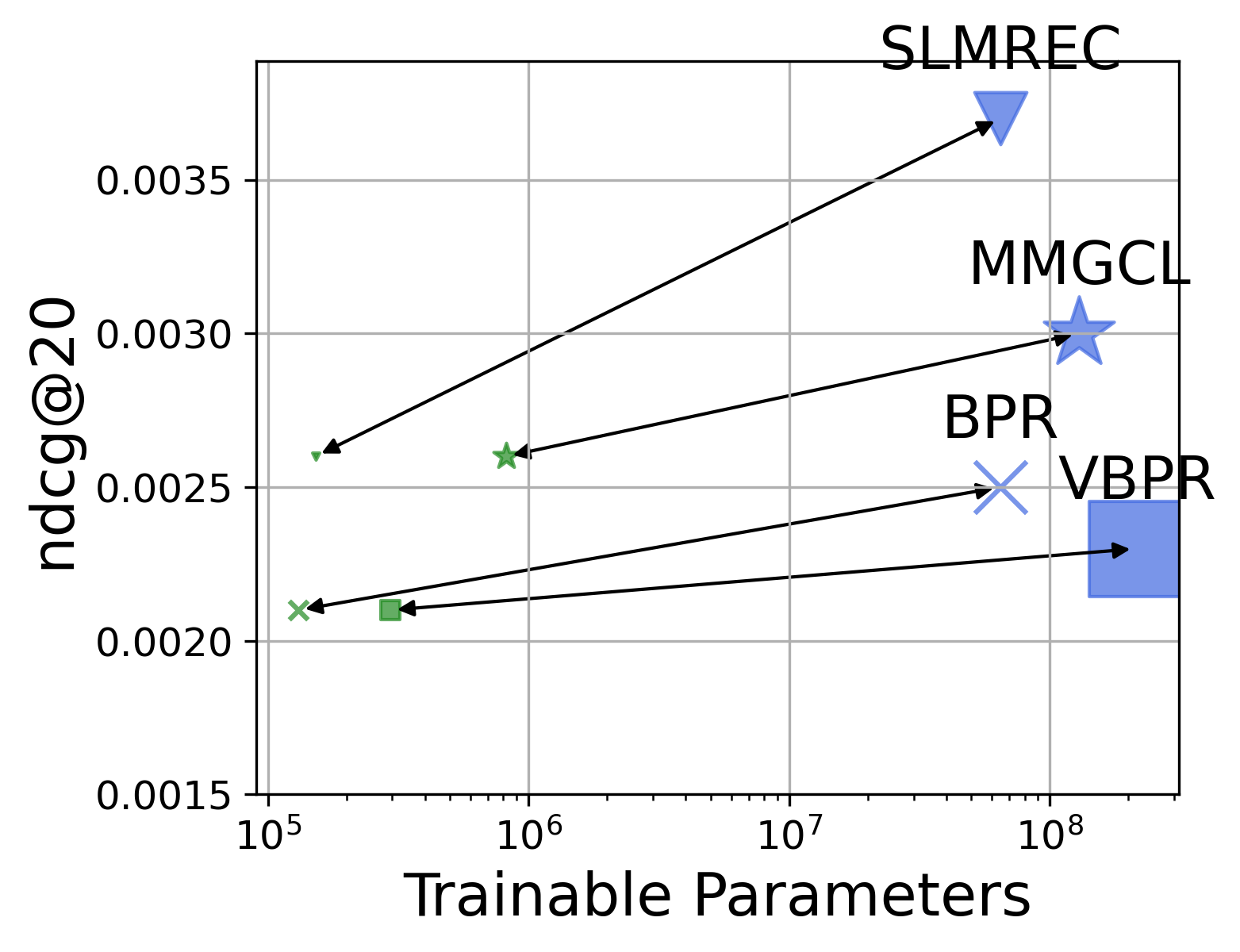}
        \caption{NDCG on Sports23}
        \label{fig:rq1_sports_ndcg}
    \end{subfigure}

    \caption{Model performance across ML1M, Baby23, and Sports23 datasets. The left column shows Recall@20 and the right column shows NDCG@20. The x-axis (log scale) reports the number of trainable parameters. Blue symbols represent the vanilla (full) models, while green symbols represent their corresponding quantized variants; arrows indicate parameter reduction. Marker size indicates the model checkpoint size. TOST confirms statistical equivalence for all three datasets.}
    \label{fig:rq1_combined}
\end{figure*}

\subsection{RQ1: Performance Analysis}

To answer RQ1, Figure~\ref{fig:rq1_combined} reports the results obtained for the ML1M, Baby23, and Sports23 datasets for both the Recall and NDCG metrics. 
From the figures, we observe for {all three} datasets a marked decrease in terms of the trainable parameter counts. {On Sports23 and Baby23}, these counts decrease by three orders of magnitude, from $10^8$ when no quantization is applied to $10^5$ when URecJPQ is applied. {On ML1M, this decrease is less pronounced, but this is due to the lower number of users and items in the datasets (see Table~\ref{tab:dataset_stats})}. 

Regarding the checkpoint sizes, depicted by the size of the points, we also observe a dramatic decrease: for example, the checkpoint size of VBPR on Baby23 is 325~MB, which decreases to 14~MB when URecJPQ is applied. Similarly, the MMGCL checkpoint size is almost 500~MB, while its quantized variant is only 15~MB.

In relation to recommendation performance, for the Baby23 dataset (Figures~\ref{fig:rq1_baby_recall} and~\ref{fig:rq1_baby_ndcg}), we observe that MMGCL and SLMRec perform slightly better than their quantized counterparts. In contrast, applying URecJPQ to BPR and VBPR leads to an improvement in Recall and NDCG. Moreover, URecJPQ-MMGCL outperforms its vanilla version in terms of NDCG.  
Similar findings can be observed for the Sports23 dataset (Figures~\ref{fig:rq1_sports_recall} and~\ref{fig:rq1_sports_ndcg}). Interestingly, in this case, VBPR achieves a lower Recall than its quantized counterpart, while the other models all slightly outperform their quantized versions. A similar trend is observed for NDCG. 

Finally, on ML1M, we observe improved performance for all models and all metrics (1 exception:  Recall for SLMRec). This is likely due to the fact that URecJPQ can act as a regularization mechanism to prevent overfitting, as observed in \cite{petrov2024recjpq}. 
{A qualitative analysis based on the plot of the resulting embeddings, reported in Section~\ref{sec:qualitative_analysis}, further confirms this observation.}
Conversely, on Sports23 (Figures~\ref{fig:rq1_sports_recall} and~\ref{fig:rq1_sports_ndcg}), as we already observed, performance slightly declines, except for quantized VBPR. We hypothesize an inverse linear relationship between dataset sparsity and performance gains: while compression regularizes denser datasets (ML1M, Baby23), it bottlenecks scarce CF signals as sparsity increases, causing the degradations seen on Sports23.

{It is worth pointing out that, although URecJPQ is effective in reducing the number of trainable parameters and the checkpoint sizes of the models without compromising the overall effectiveness of the models, as our results confirm, we do not observe substantial improvement in terms of training times. This aspect was also observed by \cite{petrov2024recjpq} in the RecJPQ paper, and it is mainly due to the fact that the full embeddings are reconstructed when they are required, thus the forward and backward passes still operate on tensors of the original dimensionality, and the computational savings are limited to the storage and update of the (much smaller) sub-embedding tables rather than to the overall training computation. We leave a more systematic investigation of training efficiency, including wall-clock time and convergence speed across different backbone models, as future work.}

To conclude for RQ1, we observe that the quantized models exhibit {better performance in the best case, and} only a marginal decrease in performance---averaging 8.5\% in Recall and 16\% in NDCG---{in the worst cases}, while achieving substantial reductions in checkpoint sizes and the number of trainable parameters, by up to 98\% and 99\%, respectively. These results show how URecJPQ can support the deployment of industrial RSs, where recommendations must be provided to a massive number of users.




\begin{figure} 
    \centering

    \begin{minipage}[c]{0.5\columnwidth}
        \centering
        \includegraphics[width=\textwidth]{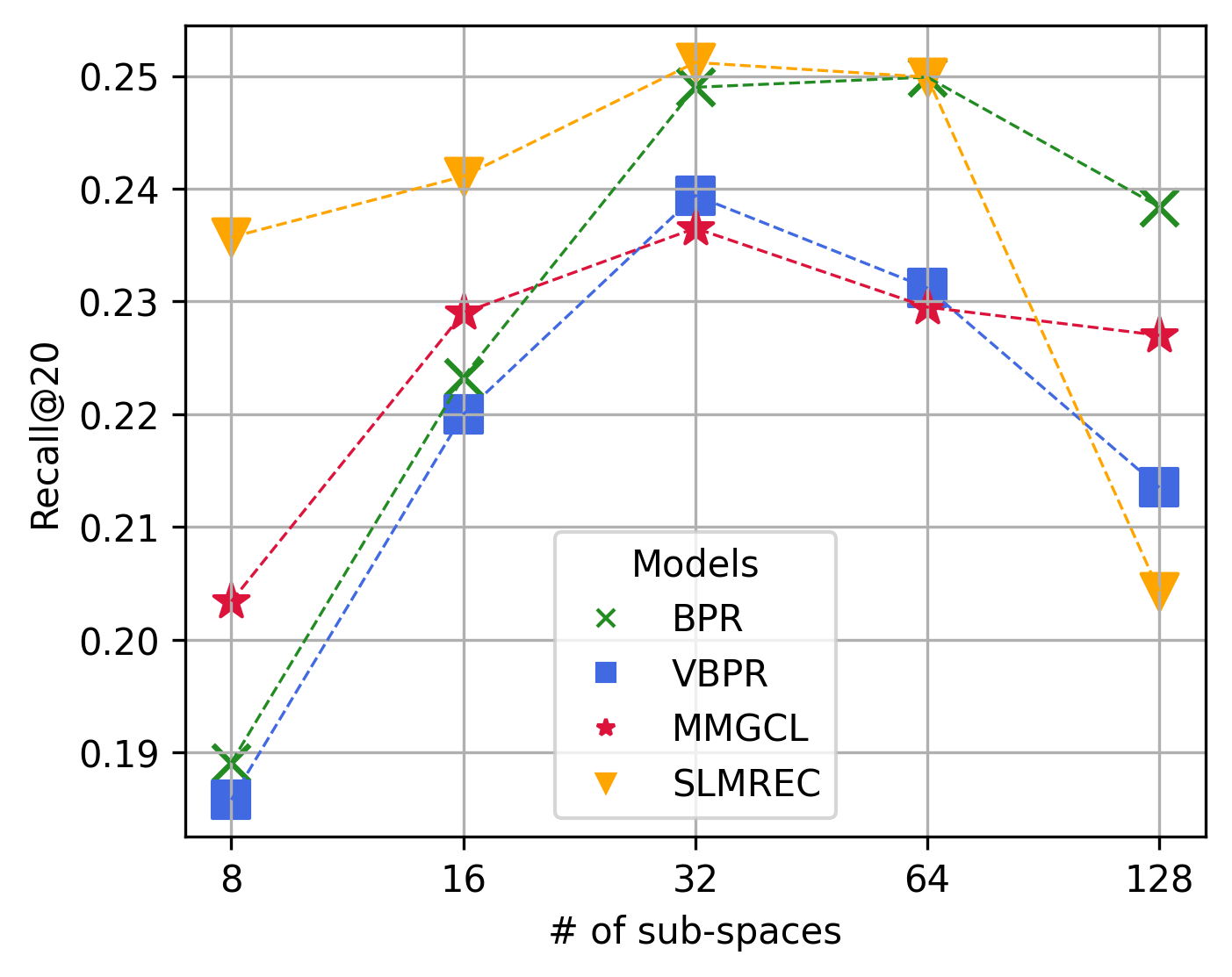}
        \small (a) ML1M
    \end{minipage}

    \vspace{1.5em} 

    \begin{minipage}[c]{0.5\columnwidth}
        \centering
        \includegraphics[width=\textwidth]{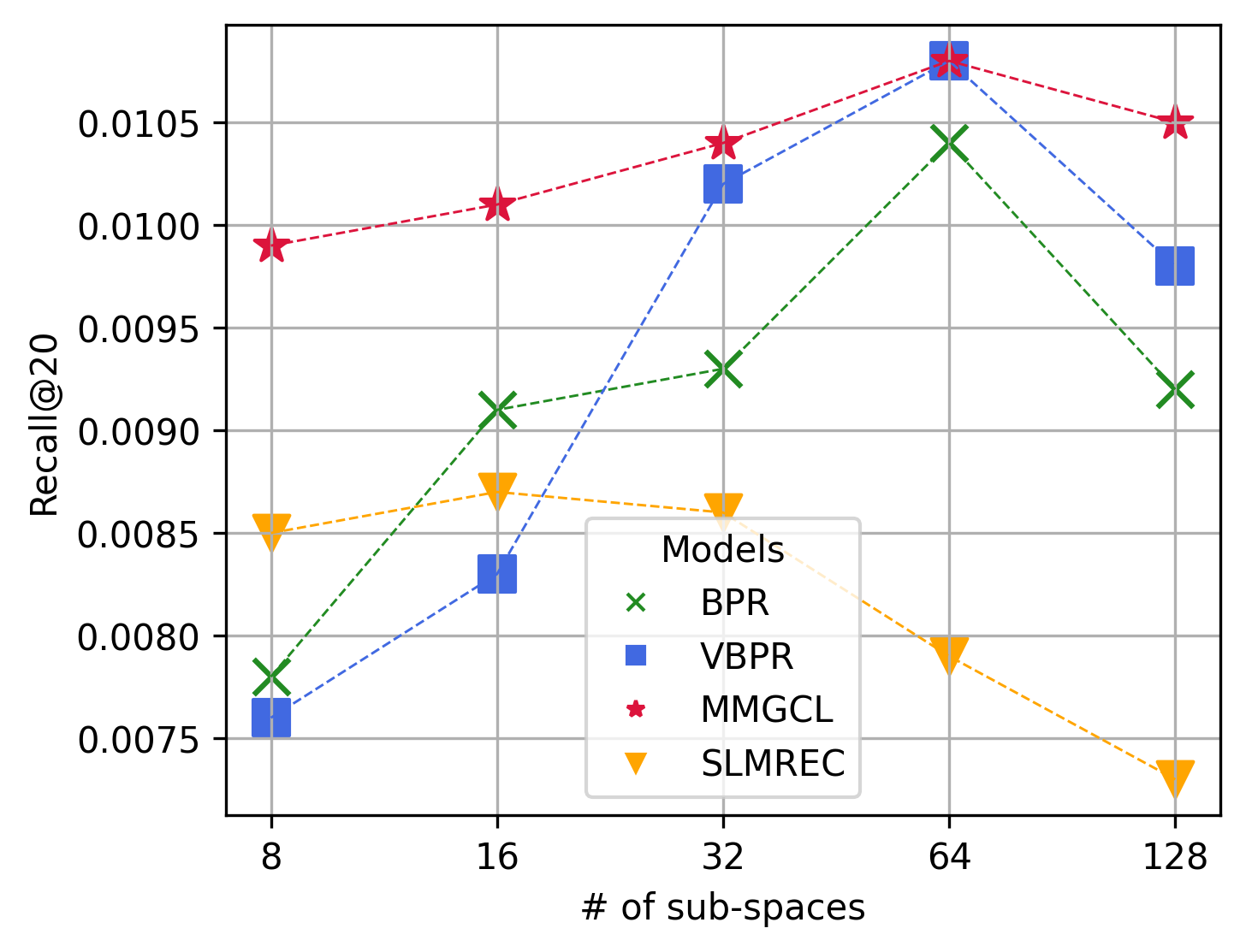}
        \small (b) Baby23
    \end{minipage}

    \vspace{1.5em}

    \begin{minipage}[c]{0.5\columnwidth}
        \centering
        \includegraphics[width=\textwidth]{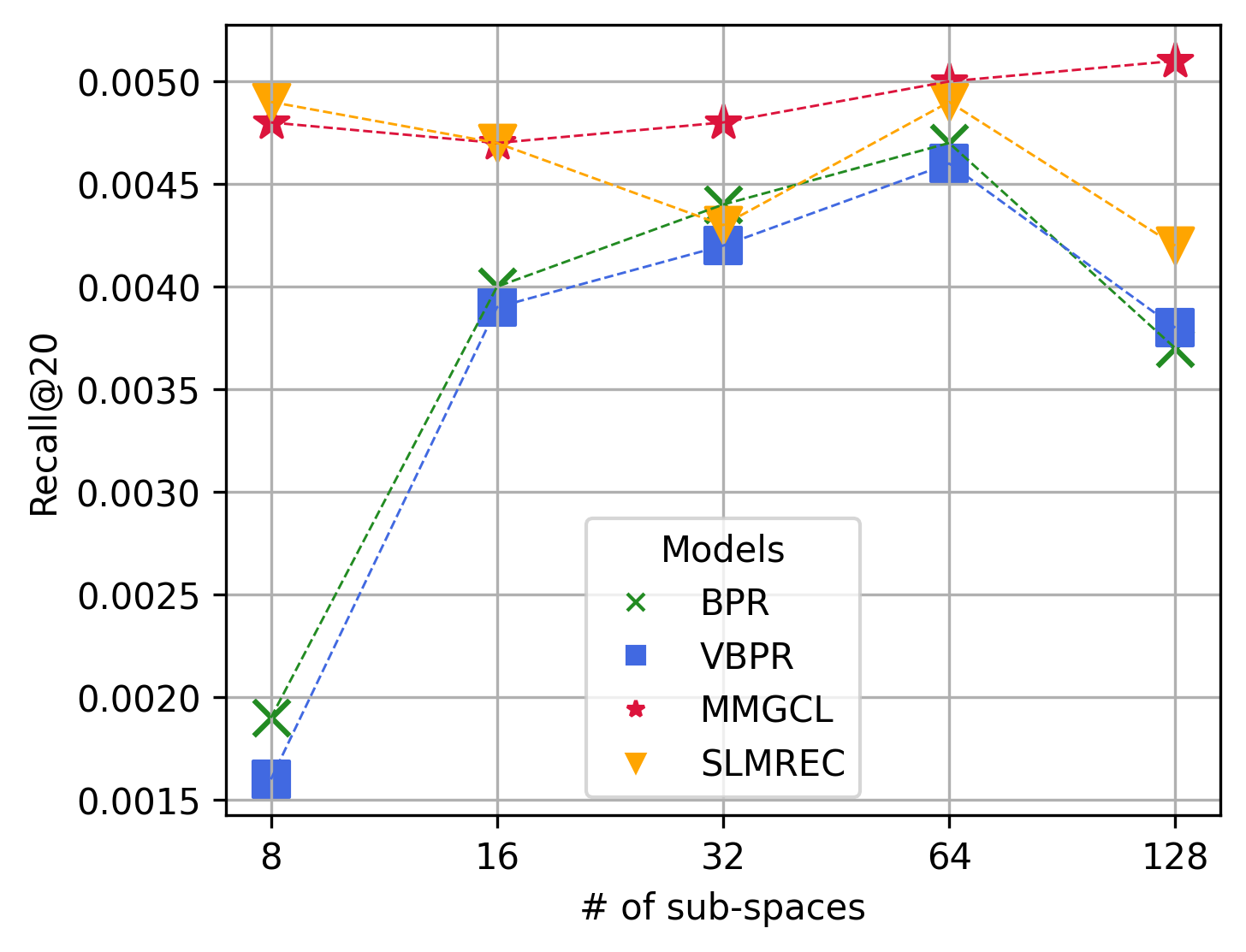}
        \small (c) Sports23
    \end{minipage}

    \vspace{1em} 
    \caption{Impact of number of sub-spaces on Recall across the different datasets.}
    \label{fig:rq2_space}
\end{figure}

\subsection{RQ2: Sub-spaces Number Sensitivity}

To answer RQ2, Figure~\ref{fig:rq2_space} reports how different numbers of sub-spaces $m$ affect recommendation performance. We focus on Recall, though similar findings were observed for NDCG. For these plots, we consider the best performance obtained by the quantized models for each number of sub-spaces. 

{We observe that quantizing with $m=32$ sub-spaces achieves the best results on ML1M, which is a small and dense dataset. The only exception is BPR, for which the best-performing number of sub-spaces is $m=64$. On the other hand, when moving to much larger and sparser datasets,} performance generally increases with the number of sub-spaces, peaking at $m=64$ before beginning to decrease. The only exceptions to this behavior are SLMRec, which drops earlier on Baby23 and shows a less monotonic trend on Sports23, and MMGCL on Sports23, which reaches its peak performance at $m=128$ sub-spaces.

To conclude for RQ2, we observe that {quantizing the embeddings using $m=32$ sub-spaces for smaller datasets and $m=64$ sub-spaces for sparser datasets} leads to the best performance in terms of Recall among the configurations considered.

\subsection{RQ3: Code Assignment Strategy Sensitivity}

\begin{table}[tb]
\centering
\resizebox{0.8\textwidth}{!}{
\renewcommand{\arraystretch}{1.2}
\begin{tabular}{lcccccc}
\hline
\multirow{2}{*}{{Model}} & \multicolumn{2}{c}{{ML1M}} & \multicolumn{2}{c}{{Baby23}} & \multicolumn{2}{c}{{Sports23}} \\
 & {Random} & {SVD} & {Random} & {SVD} & {Random} & {SVD} \\
\hline
BPR    & \textbf{0.2499} & 0.1147 & \textbf{0.0104} & 0.0056 & \textbf{0.0047} & 0.0019 \\
VBPR   & 0.1629 & \textbf{0.2394} & \textbf{0.0108} & 0.0061 & \textbf{0.0046} & 0.0017 \\
MMGCL  & 0.1961 & \textbf{0.2365} & \textbf{0.0108} & 0.0094 & 0.0048 & \textbf{0.0051} \\
SLMRec & 0.2031 & \textbf{0.2512} & 0.0073 & \textbf{0.0087} & 0.0042 & \textbf{0.0049} \\
\hline
\end{tabular}
}
\caption{Impact of code assignment strategy on performance (Recall@20).}
\label{tab:strategy_recall}
\end{table}

To address RQ3, Table~\ref{tab:strategy_recall} reports the impact of random and SVD-based code-assignment strategies on performance in terms of Recall@20. For each dataset and model, we report the best result achieved under each strategy.

First, a paired TOST on per-user Recall@20, with an equivalence bound of $\epsilon = 0.1$, indicates that the two strategies are statistically equivalent within this margin ($p < 0.05$). Nevertheless, consistent numerical differences emerge across datasets and models, suggesting that the choice of code-assignment strategy influences model behavior even when differences are not statistically significant under the chosen equivalence threshold.

On the small and dense ML1M dataset, SVD-based code assignment consistently provides higher Recall@20 for all multimodal models (VBPR, MMGCL, and SLMRec), whereas the pure collaborative filtering model, BPR, benefits substantially from the random strategy. This highlights a clear interaction between model complexity and code assignment in low-sparsity settings. 
When moving to larger and sparser datasets, the relative effectiveness of the two strategies changes. On Baby23, the random strategy achieves higher Recall in three out of four cases, while on Sports23 it performs best for two out of four models. In these settings, BPR and VBPR exhibit a higher sensitivity to the choice of code assignment, showing large performance gaps between random and SVD strategies. Notably, on Sports23, the SVD strategy is consistently suboptimal for both BPR and VBPR.

This behavior can be attributed to the combination of model simplicity and dataset sparsity. BPR and VBPR primarily rely on learning user and item ID embeddings (augmented with multimodal features in the case of VBPR). Replacing full embeddings with quantized representations substantially reduces model capacity. Under these conditions, SVD-based sub-ID assignments may fail to encode sufficiently informative user--item relationships, especially in sparse datasets, introducing noise that degrades performance. In contrast, random sub-ID assignments promote greater diversity in the learned representations, which appears beneficial for these simpler models.

For more complex architectures such as SLMRec and MMGCL, which perform higher-order modeling of interactions and multimodal signals, the impact of the code-assignment strategy is less pronounced. As shown in Table~\ref{tab:strategy_recall}, these models often exhibit smaller performance gaps between random and SVD assignments; in some cases, the SVD strategy yields better results, as observed for SLMRec on both Baby23 and Sports23 and for MMGCL on Sports23.

Finally, it is worth noting that Sports23 is considerably sparser than Baby23 (Table~\ref{tab:dataset_stats}). While this increased sparsity negatively affects SVD-based assignments for simpler models, it does not prevent more expressive models from benefiting from structured sub-ID assignments. A deeper investigation into how dataset sparsity interacts with model complexity and code-assignment strategies is left for future work.

To conclude regarding RQ3, we observe that the code-assignment strategy is a relevant factor in determining model effectiveness, with its impact strongly dependent on both model complexity and dataset characteristics. Random code assignment generally favors simpler models, such as BPR and VBPR, especially on larger and sparser datasets where it leads to significantly higher Recall. In contrast, more complex models, SLMRec and MMGCL, are less sensitive to the choice of strategy and can benefit from structured SVD-based assignments, even under high sparsity. Overall, while the two strategies are statistically equivalent within the chosen equivalence margin, the consistent numerical trends in Table~\ref{tab:strategy_recall} highlight meaningful interactions between code assignment, model architecture, and data sparsity.

\subsection{Summary}

Overall, our experimental evaluation of URecJPQ across different datasets and backbone models reveals three key insights. 

First, regarding \textbf{RQ1}, we demonstrate that URecJPQ achieves a dramatic reduction in both the number of trainable parameters (up to 99\%) and checkpoint sizes (up to 98\%) while maintaining comparable recommendation quality. In dense settings such as ML1M, the quantization framework acts as an effective regularization mechanism that often improves performance over vanilla counterparts, whereas in highly sparse settings like Sports23, we observe a slight performance trade-off due to the bottlenecking of scarce collaborative filtering signals. {The regularization mechanism is also confirmed by a qualitative analysis conducted and discussed in Section~\ref{sec:qualitative_analysis}.}

Second, answering \textbf{RQ2}, we find that the optimal number of sub-spaces $m$ is closely tied to dataset scale and density. Specifically, $m=32$ sub-spaces yield the best results for smaller, dense catalogs, while $m=64$ provides the necessary granularity for larger and sparser scenarios, with performance typically declining beyond these points due to over-segmentation of the latent space. 

Finally, for \textbf{RQ3}, we observe that the effectiveness of the code-assignment strategy is highly dependent on model complexity and architecture. Random assignment generally favors simpler, ID-heavy models like BPR and VBPR—particularly in sparse datasets where it promotes representation diversity—whereas structured SVD-based assignments are better suited for complex, multimodal architectures like SLMRec and MMGCL that can leverage higher-order interaction modeling. 

To conclude, these results validate URecJPQ as a robust and memory-efficient solution for large-scale and multimodal recommendation, enabling the deployment of state-of-the-art models in industrial, memory-constrained environments.

\section{Qualitative Analysis}\label{sec:qualitative_analysis}

\begin{figure*}
    \centering

    \begin{subfigure}[b]{0.48\textwidth}
        \centering
        \includegraphics[width=\textwidth]{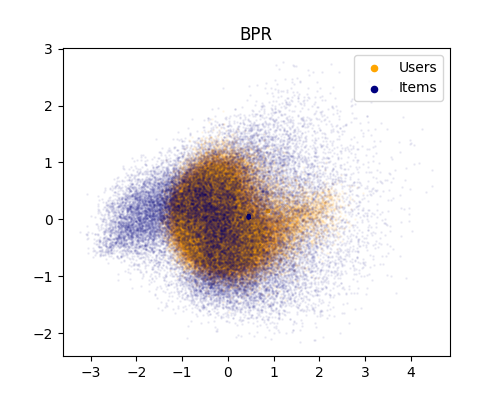}
        \label{qualitative_bpr}
    \end{subfigure}
    \begin{subfigure}[b]{0.48\textwidth}
        \centering
        \includegraphics[width=\textwidth]{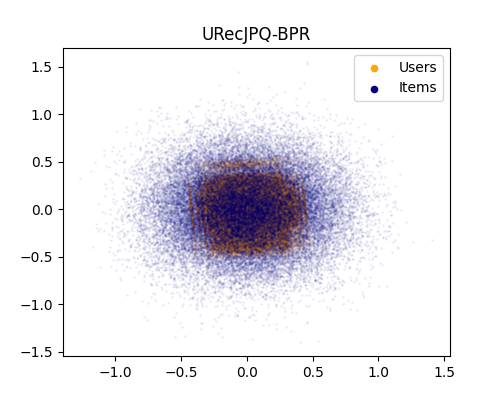}
        \label{qualitative_jpq_bpr}
    \end{subfigure}


    \begin{subfigure}[b]{0.48\textwidth}
        \centering
        \includegraphics[width=\textwidth]{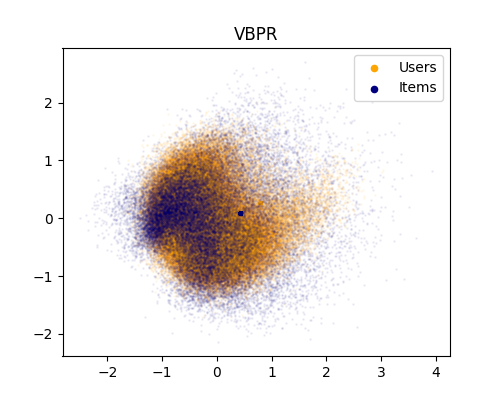}
        \label{qualitative_vbpr}
    \end{subfigure}
    \begin{subfigure}[b]{0.48\textwidth}
        \centering
        \includegraphics[width=\textwidth]{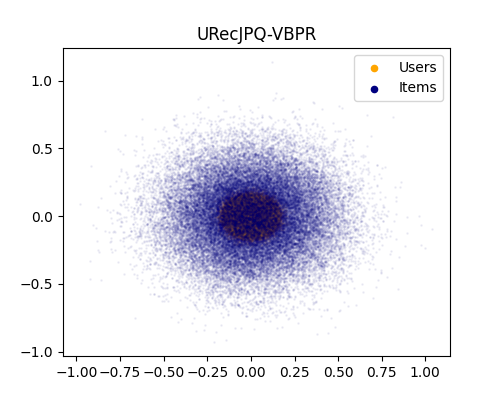}
        \label{qualitative_jpq_vbpr}
    \end{subfigure}

    \caption{PCA applied to user and item embeddings learned by BPR and VBPR, for both vanilla models (on the left) and quantized models (on the right). Yellow scatters represent users, while blue scatters represent items.}
    \label{fig:qualitative}
\end{figure*}

{To further assess understand the impact of the quantization performed by URecJPQ on the latent space, we perform a qualitative analysis by visualizing the user and item embeddings. 
Specifically, we apply Principal Component Analysis (PCA) to reduce the user and item embeddings to a 2D space for both the vanilla and quantized versions of BPR and VBPR, on the Baby23 dataset\footnote{Due to space reasons, we limit our discussion on these two models, as VBPR is the multimodal extension of BPR, thus the comparison is more direct.}. To this aim, in the case of quantized models, we first reconstruct the full embeddings by concatenating the sub-embeddings, then we apply the PCA.}

{Through this process, each embedding is represented as a scatter in a 2D space. Specifically, user embeddings are represented as yellow scatters, while item embeddings are represented as blue scatters.
The PCA plots are provided in Figure~\ref{fig:qualitative}, which reveal several key insights into the behavior of URecJPQ.}

{The vanilla BPR and VBPR models exhibit a wide, crescent-shaped distribution. In contrast, the embeddings reconstructed from URecJPQ are significantly more compact and centered around the origin. This visual density confirms that the shared sub-embedding mechanism acts as a powerful regularization tool, preventing the model from assigning excessive importance to unique IDs that may lead to overfitting.
Moreover, across the quantized plots, user embeddings (in yellow) appear more concentrated than item embeddings (in blue). This concentration suggests that URecJPQ effectively clusters users with similar interaction patterns into a shared latent core, which is particularly beneficial in the large-scale Baby23 scenario where CF signals can be scarce.
As for the performance, we already discussed that vanilla BPR and VBPR consistently perform worse than their quantized counterparts. The visual compactness of the quantized models correlates with the observed performance improvements, confirming that the bottleneck created by Joint Product Quantization helps the model capture global interaction signals more effectively than unconstrained embeddings.}

{To provide a quantitative perspective on the latent space structure, we compute the average intra-subspace distances for the learned sub-embeddings. For the BPR model, the average distance between user sub-embeddings is significantly lower than that for items ($0.16$ vs. $0.41$), a trend that persists in the VBPR model ($0.15$ vs. $0.32$). 
The smaller distances for users indicate a high degree of parameter sharing and clustering, which reinforces the hypothesis that URecJPQ acts as a robust regularization mechanism. 
By constraining users to a more compact region of the latent space, the model effectively mitigates overfitting, leading to the better performance observed in the Baby23 dataset compared to unconstrained vanilla models. 
Furthermore, the reduction in item sub-embedding distances in VBPR compared to BPR ($0.32$ vs. $0.41$) suggests that the integration of multimodal features provides a stronger external signal for item alignment, allowing the ID-based sub-embeddings to be even more compact, leading to better recommendation performance.}

{These plots confirm that URecJPQ does not merely compress the model but actively optimizes the representation space, leading to more robust recommendations in industrial-scale settings.}

\section{Conclusions}\label{sec:conclusions}
In this paper, we proposed URecJPQ, an extension of RecJPQ to the top-\textit{k} recommendation task.
The experiments conducted in a large-scale and multimodal recommendation setting showed that URecJPQ is an effective solution for dramatically reducing the number of trainable parameters (with reductions of up to 98\%), while incurring only marginal performance drops (8.5\% on average), or even {statistically equivalent} improvements when considering the BPR, VBPR and MMGCL models.
{A qualitative analysis confirmed that the quantization approach effectively acts as a regularization mechanism, preventing overfitting without losing recommendation accuracy.}
This supports the potential impact of our methodology in industrial large-scale scenarios, where our quantization approach can offer significant practical benefits, allowing training recommendation models on even larger datasets.

In future work, we plan to explore additional recommendation models, e.g., graph-based models, such as LATTICE (\cite{zhou2023tale}), FREEDOM (\cite{zhang2021mining})), {and MMGCN (\cite{wei2019mmgcn}) which would require dedicated tuning and architectural adaptations to properly handle their additional trainable parameter related to the graph structure. We also plan to consider additional} datasets to further generalize our findings.
Furthermore, we plan to investigate strategies to also compress multimodal embeddings, {and the assessment of the theoretical derivation of quantization error bounds for the joint approximation of user and item embeddings.}
Finally, we plan to investigate the quantization impact on beyond-accuracy aspects of recommendation models, including fairness, novelty and diversity of the recommendation lists, as well as the overall energy efficiency during training.

\subsection{Published Version}
This manuscript has been published in the Journal of Intelligent Information Systems: \url{https://link.springer.com/article/10.1007/s10844-026-01088-x}

\bibliography{biblio}

\end{document}